\documentclass[sn-aps]{sn-jnl}

\jyear{2021}%

\theoremstyle{thmstyleone}%
\theoremstyle{thmstyletwo}%
\theoremstyle{thmstylethree}%

\begin{document}

\title[Studying the Impact of Virtuality-Dependent Nucleon Structure $\ldots$]{Studying the Impact of Virtuality-Dependent Nucleon Structure Modification on Spectator-Tagged Deep Inelastic Scattering}

\author[1]{\fnm{Sara} \sur{Ratliff}}
\author*[1]{\fnm{Axel} \sur{Schmidt}}\email{axelschmidt@gwu.edu}

\affil[1]{\orgdiv{Department of Physics}, \orgname{The George Washington University}, \orgaddress{\city{Washington}, \state{DC} \postcode{20052}, \country{USA}}}

\abstract{Measurements of deep inelastic scattering from nuclei have revealed that the partonic structure of bound nucleons
    differs from that of free nucleons. One hypothesis is that this structure modification primarily occurs in highly virtual nucleons 
    participating in short-range correlations, although distinguishing this from other hypotheses is difficult with inclusive measurements alone. 
    Spectator-tagged deep inelastic scattering, on the other hand, may offer a way to specifically probe the partonic structure of highly-virtual
    nucleons by detecting the correlated emission of a spectator nucleon. Here, we present a method for calculating a ``spectator-tagged'' structure
    function for a nucleus by combining Generalized Contact Formalism's description of short-range correlations with light-cone convolution formalism
    to determine the impact of nucleon motion on the structure function. We apply this method to calculate predictions for helium-4, and find that
    differences in the virtuality-dependence of nucleon structure modification can lead to large measurable changes in the tagged structure function.
    The recent CLAS12 Short-Range Correlations Experiment, which collected electron scattering data on helium-4 and other nuclear targets, may be able 
    to constrain this virtuality-dependence and help test whether correlations are the origin of the modification of bound nucleon structure.}


\maketitle

\section{Introduction}

Deep inelastic scattering (DIS) measurements on nuclei have revealed that the partonic structure of nuclei differs from that of
free nucleons in substantial ways~\cite{EuropeanMuon:1983wih,Arnold:1983mw,BCDMS:1987upi,Gomez:1993ri,Seely:2009gt}. Coherent destructive and constructive interference effects, dubbed ``shadowing'' and
``anti-shadowing'' respectively, alter the nuclear $F_2$ structure function below the valence region. In the valence
region, a universal reduction in $F_2$ has been observed, which is now called the ``EMC Effect.'' The EMC Effect grows
with nuclear size, saturating at about a 15\% reduction at a momentum fraction of approximately $0.7$ in the heaviest
nuclei. While conventional nuclear physics effects such as Fermi motion and nuclear pions play a role, the magnitude
of the effect requires a modification of the intrinsic structure of bound nucleons. The underlying mechanism driving
the effect is not conclusively known. See Refs.~\cite{Malace:2014uea,Hen:2016kwk} for recent
reviews of the topic.

One of the many hypotheses for the cause of the EMC Effect is that the observed reduction in $F_2$ is driven primarily by a
large modification to the structure of nucleons participating in a short-range nucleon-nucleon correlations. Short-range correlations (SRCs)
consist of pairs of nucleons in close proximity, whose strong interactions lead to large relative momenta, often well above the 
nuclear Fermi momentum. As a result, correlated nucleons tend to be the highest virtuality nucleons in the nucleus.
Like the EMC Effect, such correlations appear to be a universal feature of nuclear structure, and have been experimentally observed in 
a wide range of nuclei through a wide range of experimental techniques (e.g., \cite{Piasetzky:2006ai,Fomin:2011ng,Terashima:2018bwq}). This universality is understood to emerge from the short-range effective nucleon-nucleon interaction~\cite{Frankfurt:1993sp,CiofidegliAtti:1995qe}. See Ref.~\cite{Arrington:2022sov} for more details on recent experimental progress.

The EMC-SRC hypothesis is supported by several experimental and theoretical findings. There is a striking correlation between the magnitude of the EMC Effect and the prevalence of SRC pairs across all measured nuclei~\cite{Weinstein:2010rt,Hen:2012fm,Arrington:2012ax}. Effective field theory calculations suggest that this correlation is a natural consequence of scale separation~\cite{Chen:2016bde}. Schmookler et al.\ have shown that nuclear DIS cross section ratios are consistent with the universal modification of the nucleons in SRCs~\cite{CLAS:2019vsb}. Universal modification would have consequences for extractions of free neutron structure from nuclear DIS data~\cite{Segarra:2019gbp}, and these predictions are largely consistent with recent results from the MARATHON Experiment~\cite{JeffersonLabHallATritium:2021usd}.

There are multiple possibilities for the mechanism driving structure modification of SRCs. Due to their close proximity, nucleons in short-range correlations experience significantly higher nuclear density than typical nucleons. This may even allow quarks in the two nucleons to interact, an idea considered in six-quark bag~\cite{Bickerstaff:1984gut,CiofidegliAtti:1994vf} and hidden color models~\cite{West:2020rlk,West:2020tyo}. Alternatively, correlated nucleons have momenta exceeding the nuclear Fermi momentum and are highly virtual, which could lead to modification~\cite{CiofidegliAtti:2007ork}. Arrington and Fomin have attempted to study whether local-density dependence or high-virtuality dependence are more consistent with experimental data, without strong conclusions~\cite{Arrington:2012ax,Arrington:2019wky}. An analysis of quantum Monte Carlo calculations indicates that there may not be a distinction between local-density and virtuality dependences~\cite{Cruz-Torres:2019fum}. 
There are also studies which challenge the hypothesis that SRCs are the source of the modification observed in the EMC 
effect~\cite{Paris:2000bj,Wang:2020uhj}. Given the range of theoretical perspectives, new experimental approaches would be helpful in clarifying the role SRCs play in the EMC Effect.

Efforts to understand the EMC Effect and the EMC-SRC hypothesis eventually run into the limits of what can be learned by DIS data alone. Inclusive nuclear DIS measurements are sensitive to the average modification of all nucleons. Efforts have been made to study how the EMC Effect changes across light nuclei with different nuclear structures~\cite{Seely:2009gt,Arrington:2021vuu}, but these systematic efforts will always face difficulties in establishing unambiguous conclusions. New experimental techniques that add additional information beyond what inclusive DIS can provide are needed.

A new experimental technique which holds promise for testing the EMC-SRC hypothesis is spectator-tagged DIS, in which a spectator nucleon is detected in coincidence with the outgoing electron from DIS. 
The detection of a spectator, specifically one with high momentum oriented anti-parallel to the momentum transfer, is a strong indication that the electron scattered from a nucleon in an SRC. Spectator-tagged DIS was measured on deuterium at Jefferson Lab in the 6-GeV era~\cite{CLAS:2005ekq,CLAS:2011qvj,CLAS:2014jvt}, and two dedicated experiments, BAND~\cite{BAND:proposal} and LAD~\cite{LAD:proposal} were proposed to study the relationship between SRCs and structure modification in deuterium. At the time of this writing, BAND has collected data and is under analysis, while LAD is expected to run in a little more than one year. In addition, the BAND Experiment necessitated the construction of the BAND detector~\cite{Segarra:2020txy}, a Backward Angle Neutron Detector for detecting spectator neutrons that is now part of the standard equipment of the CLAS-12 Spectrometer in Jefferson Lab Hall B~\cite{Burkert:2020akg}. Spectator neutron-tagging analyses can be performed on data from a wide-variety of CLAS-12 experiments.

BAND and LAD targeted deuterium because the system of two nucleons is simpler to study, even though any structure modification effects are expected to be smaller. In this paper, we consider the possibility of studying spectator-tagged DIS on $^4$He; a slightly more complicated system, but one whose EMC Effect is larger, and on which abundant data have been collected as part of a recent CLAS-12 Experiment~\cite{RGM:proposal}. We describe a formalism for estimating the spectator-tagged spectral function for a nucleus, building on the convolution framework developed in Ref.~\cite{Segarra:2020plg}, and modeling short-range correlations using the light-cone formulation of Generalized Contact Formalism, described in Ref.~\cite{Pybus:2020itv}. We perform calculations under several different assumptions for the virtuality-dependent modification. We find that a reasonable range of assumptions, which are all consistent with current data, can lead to a wide variability on the tagged-structure function for $^{4}$He. This suggests that spectator-tagging measurements on $^4$He might shed light on the dependence of structure modification in bound nucleus on virtuality, and on the role short-range correlations play in the EMC Effect.

\section{Formalism}

Here, we establish a prescription for calculating a ``spectator neutron-tagged $F_2$ structure function,'' i.e., the $F_2$ structure function for deep inelastic electron scattering from a nucleus, in which there is also the simultaneous emission of a spectator neutron. 
This prescription extends formalisms described in Refs.~\cite{Segarra:2020plg,Pybus:2020itv}. 
In particular, we employ a convolution formalism with virtuality-dependent modification developed in Ref.~\cite{Segarra:2020plg} to define a structure function for scattering from moving, off-shell nucleons with modified partonic structure. 
We use the light-cone formulation of Generalized Contact Formalism (GCF)~\cite{Pybus:2020itv} to describe the motion of nucleons along with correlated spectators participating in short-range correlations. In this section, we detail how these formalisms can be combined to allow a calculation of the spectator-tagged structure function.

For what follows, we consider deep inelastic electron scattering, in which the scattered electron transfers energy $\nu$ and momentum $\vec{q}$ to a single parton in a nucleus with $Z$ protons, and $N=A-Z$ neutrons. We define $Q^2= \lvert \vec{q}\rvert^2 - \nu^2$ as the negative squared four-momentum transfer.
We use the Bjorken $x_B$ parameter, which is given by $Q^2/2m_N \nu$ in the fixed-target frame, where $m_N$ is the nucleon mass. We will work in a light-cone coordinate system in which momentum four-vectors are defined by $p^+=E + p_z$, $p^-=E-p_z$, and transverse components $\vec{p}^\perp = \langle p_x, p_y \rangle$. We choose the $z$ axis to point in the direction of $-\vec{q}$, i.e., in the direction opposite the 3-momentum transferred by the scattered electron. We use the term light-cone momentum fraction, denoted by the symbol $\alpha$ to be the fraction of $p^+$ momentum. Our goal is therefore to build a prescription for the spectator-tagged structure function $F_2^{A,\text{tag}}(x_B,Q^2,\alpha_2,\vec{p}_2^\perp)$, which, in addition to the familiar dependence on $x_B$ and $Q^2$, depends on the light-cone momentum fraction of the spectator, $\alpha_2$, as well as its transverse momentum $\vec{p}_2^\perp$. We will use the index `2' to refer to the spectator nucleon, and index `1' to refer to the struck nucleon.

In convolution formalism, the inclusive structure function for scattering from a nucleus, $F_2^A(x_B,Q^2)$ can be defined in terms of light-cone
density functions, $\rho_{1}(\alpha_1,v_1)$, which describe the distribution for nucleons in the nucleus to have light-cone momentum
fraction $\alpha_1$ and virtuality $v_1 \equiv (E_1^2 - p_1^2 - m_N^2)/m_N^2$~\cite{Segarra:2020plg}. In terms of a light-cone density for protons, $\rho_{1,p}(\alpha_1,v_1)$, and $\rho_{1,n}(\alpha_1,v_1)$ for neutrons, the inclusive structure function is given by:
\begin{multline} 
F_2^A(x_B,Q^2) = \frac{1}{A}\int_{x_B}^A \frac{d\alpha_1}{\alpha_1}
\int_{-\infty}^0 dv_1 \\
\left[
Z \rho_{1,p}(\alpha_1,v_1)F_2^p(\tilde{x},Q^2) + N\rho_{1,n}(\alpha_1,v_1)F_2^n(\tilde{x},Q^2)
\right],
\end{multline}
where $F_2^p$ is the free proton structure function, $F_2^n$ is the free neutron structure function, and
$\tilde{x} \equiv \frac{Q^2}{2q_\mu p_1^\mu}$ is the fraction of the off-shell nucleon's momentum carried by the struck parton.
In the Bjorken limit, $\tilde{x}\rightarrow \frac{x_Bm_N}{\alpha_1 \bar{m}}$, which was used in Ref.~\cite{Segarra:2020plg}, and
which we will adopt here. This structure function prescription can be extended to
include a virtuality-dependent modification term to get:
\begin{multline} 
F_2^A(x_B,Q^2) = \frac{1}{A}\int_{x_B}^A \frac{d\alpha_1}{\alpha_1}
\int_{-\infty}^0 dv_1 \\
\left[
Z \rho_{1,p}(\alpha_1,v_1)F_2^p(\tilde{x},Q^2) + N\rho_{1,n}(\alpha_1,v_1)F_2^n(\tilde{x},Q^2)
\right]\\
\times \left(1 + v_1 f_\text{off}(\tilde{x})\right).
\end{multline}
where $f_\text{off}(\tilde{x})$ describes the $\tilde{x}$-dependence of this modification. Ref.~\cite{Segarra:2020plg} studied the effects of using different assumptions for $f_\text{off}(\tilde{x})$, including constant and linear models.

To determine the spectator-tagged structure function, we assume that the emission of a spectator will occur when the struck nucleon and spectator nucleon are in a short-range correlation together, and can thus be described using GCF. In Ref.~\cite{Pybus:2020itv}, GCF is used to define the light-cone decay function, i.e., the distribution of pairs of correlated nucleons in the nucleus, given by
\begin{equation}
\rho(\alpha_1,\vec{p}_1^\perp, \alpha_2, \vec{p}_2^\perp) = 
\frac{\alpha_1}{\alpha_\text{CM}} 
\rho_\text{SRC}(\alpha_\text{rel.},\vec{p}_\text{rel.}^\perp) 
\rho_\text{CM}(\alpha_\text{CM},\vec{p}_\text{CM}^\perp),
\end{equation}
where $\alpha_1$,$\vec{p}_1^\perp$ and $\alpha_2$,$\vec{p}_2^\perp$ combine according to
\begin{align}
\alpha_\text{CM} &= \alpha_1+\alpha_2 \\
\alpha_\text{rel.} &= \frac{2\alpha_2}{\alpha_\text{CM}}\\
\vec{p}_\text{CM}^\perp &= \vec{p}_1^\perp + \vec{p}_2^\perp\\
\vec{p}_\text{rel.}^\perp &= \vec{p}_2^\perp - \frac{\alpha_2}{\alpha_\text{CM}}\vec{p}_{CM}^\perp \\
&= \frac{\alpha_1 \vec{p}_2^\perp - \alpha_2 \vec{p}_1^\perp}{\alpha_\text{CM}},
\end{align}
and $\rho_\text{SRC}$ and $\rho_\text{CM}$ refer to the relative and center-of-mass distributions, respectively, of nucleons within SRC pairs. Following Ref.~\cite{Pybus:2020itv}, we model $\rho_\text{CM}$ as a Gaussian:
\begin{equation}
\label{eq:rhoCM}
\rho_\text{CM}(\alpha_\text{CM},\vec{p}_\text{CM}^\perp) =
\frac{\bar{m}\alpha_\text{CM}}{(2\pi\sigma_\text{CM})^{3/2}}
\exp \left\{
-\frac{\bar{m}^2 (2-\alpha_\text{CM})^2 + \lvert \vec{p}_\text{CM}^\perp \rvert ^2}{2\sigma_\text{CM}^2}
\right\}
\end{equation}
where $\bar{m}\equiv \frac{m_A}{A}$ and $\sigma_\text{CM}$ represents the width of the CM momentum distribution, $\approx 100$--$150$~MeV$/c$~\cite{Colle:2013nna,LabHallA:2014wqo,CLAS:2018qpc}.

In Generalized Contact Formalism, the relative distribution takes the form of sum over pairs of different relative quantum numbers, which we represent by index $j$:
\begin{equation}
\label{eq:rhoSRC}
\rho_\text{SRC}(\alpha_\text{rel},\vec{p}_\text{rel}^\perp) = 
\sum_j C_j \frac{\sqrt{m_N^2+k^2}}{2-\alpha_\text{rel}} 
\frac{\lvert \tilde{\phi}_j (k)\rvert ^2}{(2\pi)^3},
\end{equation}
where $\tilde{\phi}_j(k)$ is a universal (i.e., nucleus-independent) two-body momentum space wave function, $C_j$ is the nucleus-dependent contact term representing the abundance of SRC pairs with quantum numbers $j$, and $k$ is given by:
\begin{equation}
k^2 \equiv \frac{m_N^2 + \lvert \vec{p}_\text{rel}^\perp\rvert ^2}{\alpha_\text{rel} (2-\alpha_\text{rel})} - m_N^2.
\end{equation}
The decay function can be restricted based on the isospin projection of nucleons 1 and 2, which determines which quantum numbers are considered in the sum over $j$ in Eq.~\ref{eq:rhoSRC}. 
We'll use the notation $\rho_{pn}$ to refer to the decay function where nucleon 1 is a proton and nucleon 2 is a neutron (either spin 0 or spin 1).
We reserve the notation $\rho_{1,p}$ and $\rho_{1,n}$ to refer to the decay function where nucleon 1 is a proton or neutron respectively (with no conditions on nucleon 2).

Ref.~\cite{Segarra:2020plg} gives the inclusive $F_2$ structure function for a nucleus in terms of the decay function, by integrating the decay function over the spectator nucleon variables (and re-defining $\vec{p}_1^\perp$ in terms of virtuality, $v_1$) in order to get the light-cone density:
\begin{multline}
\rho_1(\alpha_1,v_1) \equiv 
\int d^2 \vec{p}_1^\perp \int d^2 \vec{p}_2^\perp 
\int_0^A \frac{d\alpha_2}{\alpha_2} 
\rho(\alpha_1,\vec{p}_1^\perp, \alpha_2, \vec{p}_2^\perp) \\
\times \delta\left(v_1 - \frac{\bar{m}p^-\alpha_1 - \lvert \vec{p}_1^\perp\rvert ^2 - m_N^2}{m_N^2}\right)
\end{multline}
with 
\begin{equation}
p^- \equiv m_A - \frac{m_N^2 + \lvert \vec{p}_2^\perp \rvert ^2 }{\bar{m} \alpha_2} - \frac{m_{A-2}^2 + \lvert \vec{p}_\text{CM}^\perp \rvert^2}{\bar{m}(A-\alpha_1 -\alpha_2)}.
\end{equation}

In this work, we want to predict the \emph{neutron spectator-tagged} structure function, rather than the inclusive structure function. We therefore do not integrate out the spectator variables. Instead we arrive at an expression
\begin{multline}
    F_2^{A,\text{tag}}(x_B,Q^2,\alpha_2, \vec{p}_2^\perp) = 
    \frac{1}{A\alpha_2}\int d^2\vec{p}_1^\perp \int_{x_B}^A \frac{d\alpha_1}{\alpha_1} \\
    \left[
    ZF_2^p(\tilde{x},Q^2) \rho_{pn}(\alpha_1,\vec{p}_1^\perp, \alpha_2, \vec{p}_2^\perp) + NF_2^n(\tilde{x},Q^2)\rho_{nn}(\alpha_1,\vec{p}_1^\perp, \alpha_2, \vec{p}_2^\perp)
    \right] \\
    \times \left(1 + v_1 f_\text{off}(\tilde{x})\right),
\end{multline}
where $v_1$ takes a fixed value of $\frac{\bar{m}p^-\alpha_1 - \lvert \vec{p}_1^\perp\rvert ^2 - m_N^2}{m_N^2}$. We further factor out the well-constrained $F_2^p$, and assume that the ratio $F_2^n/F_2^p$ has negligible dependence on $Q^2$:
\begin{multline}
    F_2^{A,\text{tag}}(x_B,Q^2,\alpha_2, \vec{p}_2^\perp) = 
    \frac{1}{A\alpha_2}\int d^2\vec{p}_1^\perp \int_{x_B}^A \frac{d\alpha_1}{\alpha_1} F_2^p(\tilde{x},Q^2) \\
    \left[
    Z\rho_{pn}(\alpha_1,\vec{p}_1^\perp, \alpha_2, \vec{p}_2^\perp) + N\frac{F_2^n}{F_2^p}(\tilde{x})\rho_{nn}(\alpha_1,\vec{p}_1^\perp, \alpha_2, \vec{p}_2^\perp)
    \right] \\
    \times \left(1 + v_1 f_\text{off}(\tilde{x})\right).
    \label{eq:f2Atag}
\end{multline}

A number of inputs are needed both for the convolution model and for the GCF description of the light-cone decay function, and we have made some specific choices for the calculations in this work. 
First, for the free proton structure function, $F_2^p$, we have employed a parametrization developed by the HERMES Collaboration~\cite{HERMES:2011yno}, produced from a global fit to to DIS data, including some in Jefferson Lab kinematics,
using the ALLM model (described in \cite{Abramowicz:1991xz, Abramowicz:1997ms}) .
Second, for the ratio of free neutron to free proton structure functions, we have assumed the approximate form 
\begin{equation}
    \frac{F_2^n}{F_2^p}=a(1-x)^b+c,
\end{equation}
with parameter values $a=0.57,~b=2.2,~c=0.42.$, used by Segarra et al.~\cite{Segarra:2020plg}, which was a simplified description of the results of the analysis described in Ref.~\cite{Segarra:2019gbp}. This form is consistent with the results of the BoNUS~\cite{CLAS:2011qvj,CLAS:2014jvt} and MARATHON~\cite{JeffersonLabHallATritium:2021usd} experiments. 

Generalized Contact Formalism requires inputs for each of its factorized components. To describe the relative motion between correlated nucleons (Eq.~\ref{eq:rhoSRC}), $\rho_\text{SRC}$ requires universal two-body momentum distributions $ \lvert \tilde{\phi}_j (k) \rvert^2$ from a model of the nucleon-nucleon potential. We have chosen to use the AV18 potential~\cite{Wiringa:1994wb}. For the center-of-mass motion distribution of a correlated pair, given by $\rho_\text{CM}$ (Eq.~\ref{eq:rhoCM}), we assume a Gaussian distribution (as in Refs.~\cite{Pybus:2020itv,Weiss:2018tbu,CLAS:2020mom,CLAS:2020rue} and others) with a width $\sigma=100$~MeV$/c$, based on the measurement presented in Ref.~\cite{LabHallA:2014wqo}. Lastly, GCF requires values for the contact coefficients, $C_j$. We use the values determined in Ref.~\cite{Cruz-Torres:2019fum} from a momentum-space analysis of variational Monte Carlo calculations performed with the AV18 potential.

\begin{figure}[htpb!]
    \centering
    \includegraphics[width=3.5 in]{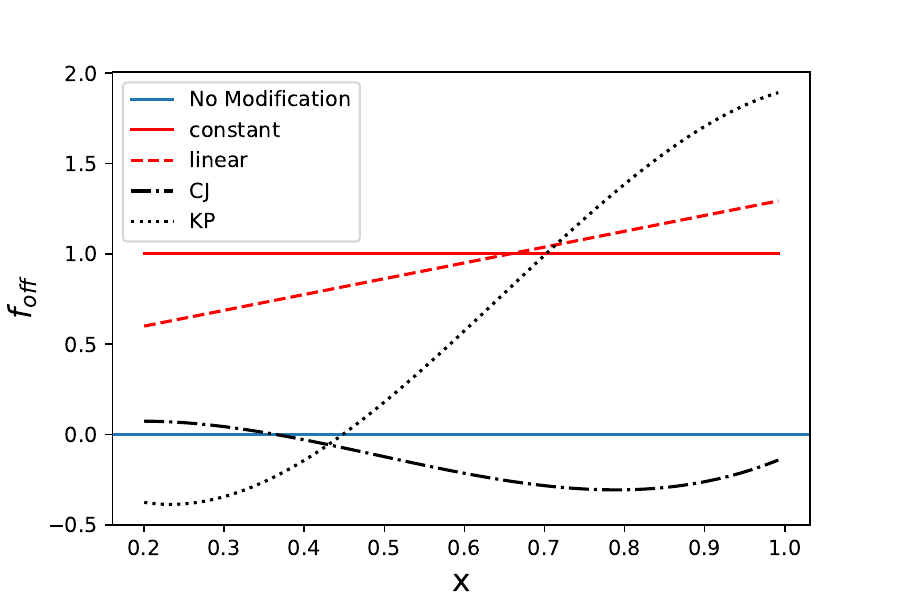}
    \caption{The different models used in this calculation for the universal off-shell nucleon modification function, $f_\text{off}$ are shown against $x$.}
    \label{fig:foff}
\end{figure}

In our study, we considered the impacts of different assumptions for the final input, the off-shell nucleon modification function, $f_\text{off}(\tilde{x})$. This function is not well-constrained by available data, and therefore we consider five possibilities
\begin{enumerate}
    \item No modification: $f_\text{off}(\tilde{x})=0$. This is inconsistent with the EMC Effect, but we use it as a useful baseline.
    \item Constant in $\tilde{x}$: $f_\text{off}(\tilde{x})=K$. We specifically take the value of $K=0.96$ from the global fit analysis of \cite{Segarra:2020plg} using the identical assumption for the free-neutron structure function.
    \item Linear in $\tilde{x}$: $f_\text{off}(\tilde{x})=A +B\tilde{x}$. Similarly, we take the value of $A=0.31$ and $B=0.98$ from the global fit analysis of \cite{Segarra:2020plg} using the identical assumption for the free-neutron structure function. 
    \item The result of Kulagin and Petti~\cite{Kulagin:2004ie} (KP), who assumed a third-order polynomial of the form
    \begin{equation}
        f_\text{off}^\text{KP}(\tilde{x})=C(x_0-\tilde{x})(x_1-\tilde{x})(1+x_0 - \tilde{x})
        \label{eq:kp}
    \end{equation}
    with parameters $C$, $x_0$, and $x_1$. We use their values: $C=8.1$, $x_0=0.448$, and $x_1=0.05$.
\item The result of the CTEQ-Jefferson Lab (CJ) collaboration~\cite{Accardi:2016qay}, which used the same functional form as in Eq.~\ref{eq:kp}, but with parameter values: $C=-3.6735$, $x_0=0.057717$, and $x_1=0.36419$.
\end{enumerate}
A comparison of these five possible $f_\text{off}$ functions is shown in Fig.~\ref{fig:foff}.  

Using these inputs, we calculated the tagged structure function $F_2^{A,\text{tag}}$ for Helium-4 by performing the integration in Eq.~\ref{eq:f2Atag} over $\vec{p}^\perp_1$ and $\alpha_1$ using Monte Carlo integration.

\section{CLAS12 Short-Range Correlations Experiment}

One motivation for performing this calculation is the opportunity presented by the CLAS12 Short-Range Correlations (CLAS12-SRC) Experiment (E12-17-006A)~\cite{RGM:proposal}, recently conducted at Jefferson Lab. The CLAS12-SRC Experiment collected data from November, 2021, through February, 2022 in Experimental Hall B, as part of Run Group M. Out-going particles were detected using the CLAS12 Spectrometer~\cite{Burkert:2020akg}, a large acceptance magnetic spectrometer designed to cover scattering angles roughly from $8^\circ$--$40^\circ$ in a toroidal forward spectrometer, and from approximately $35^\circ$--$125^\circ$ in a solenoidal central detector. In the experiment, an electron beam was scattered from a variety of nuclear targets, with the goal of identifying quasi-elastic knock-out of nucleons in SRC pairs. Several beam energies were used, with the majority of the data being collected with a 6 GeV beam. Since the experiment had a non-specific trigger---only requiring the detection of a single high-energy electron---a wide range of final states can be analyzed in the data. The targets used in the CLAS12-SRC Experiment and the time spent on each are shown in Table \ref{table:RGM Targets}. At the time of this writing, data from the experiment are undergoing calibrations and analysis. 

\begin{table}[htbp]
\centering
\caption{The targets used in the CLAS12-SRC Experiment at a beam energy of 6 GeV, the days spent on each target, and the number of triggers per target.}
\begin{tabular}{c c c} 
 \hline
 \hline
 Target & Days & Triggers ($\times 10^9$) \\
 \hline
 H & 2 & 1.1  \\ 
 D & 6 & 4.2  \\
 $^4$He & 10 & 3.9 \\
 $^{12}$C & 8 & 4.1 \\
  $^{40}$Ar & 2 & 0.55 \\
 $^{40}$Ca & 6 & 3.0\\ 
 $^{48}$Ca & 3 & 1.50 \\ 
 $^{120}$Sn & 4 & 0.40 \\
 \hline
\end{tabular}
\label{table:RGM Targets}
\end{table}

In addition to the forward and central spectrometers, the CLAS12-SRC Experiment also employed the CLAS12 Backward Angle Neutron Detector (BAND)~\cite{Segarra:2020txy}, covering 155$^\circ$--175$^\circ$ from the beamline. BAND is an array of plastic scintillator bars, read out on both ends with Photomultiplier Tubes, with a total thickness of 35 cm. BAND includes a thin layer of lead for attenuating photons, as well as an additional thin layer of active scintillator to serve as a charged particle veto. BAND was built for the BAND Experiment~\cite{BAND:proposal} (Jefferson Lab, E12-11-003A), aiming to measure spectator-tagged DIS on deuterium, with the goal of detecting neutrons in the momentum range of $\approx 200--700$~MeV$/c$. BAND was placed upstream of the target, covering backward angles, both to avoid material obstructions and because backward-spectator kinematics are less sensitive to final-state interactions~\cite{CLAS:2005ekq,Strikman:2017koc}. 

The BAND detector allows the study of spectator neutron-tagged DIS on nuclei using the CLAS12-SRC data. There are advantages and disadvantages with studying tagged DIS in nuclei as opposed to deuterium. In deuterium, there are no other nucleons outside of the interacting nucleon-nucleon pair. In the plane-wave limit, the momentum of spectator is exactly opposite to the initial momentum of the struck nucleon. In larger nuclei, the residual $A-2$ system can carry away energy and momentum. On the other hand, the magnitude of the EMC Effect grows with nuclear size. It may be easier to study the relationship between the EMC Effect and SRCs due to the larger effect sizes. Lastly, final-state interaction effects should grow with nuclear size, making the results from very large nuclei harder to interpret. 

In this paper, we specifically consider the EMC Effect helium-4, which has the advantage of being relatively light. Of the nuclear targets used in the CLAS12-SRC Experiment, it also had the longest run duration, with almost 4 billion triggers collected. The methodology of this work could easily be applied to make predictions for the other target nuclei used in the experiment.

\section{Results}

\begin{figure}[htpb!]
    \centering
    \includegraphics[width=4.8 in]{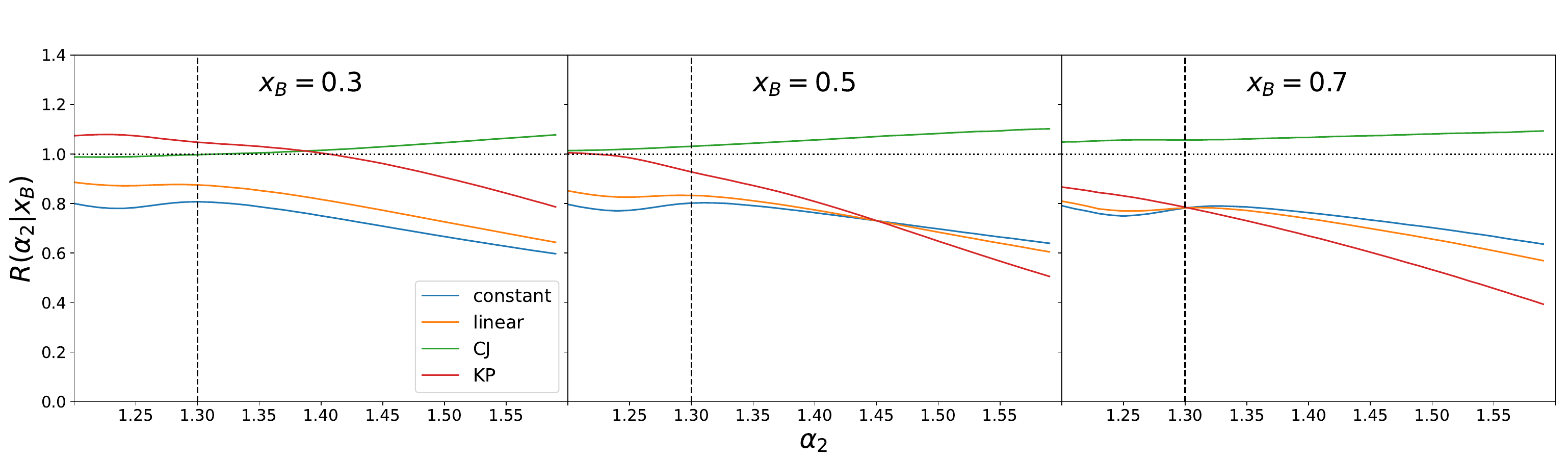}
    \caption{The results for the ratio of $F_2^A$ structure functions modified by the specified off-shell nucleon modification function models to the unmodified structure functions, shown against $\alpha_2$. Behavior to the left of the dashed line at $\alpha_2=1.3$ can be dismissed due to GCF's limitations.}
    \label{fig:single_alphadep}
\end{figure}

\begin{figure}[htpb!]
    \centering
    \includegraphics[width=4.8 in]{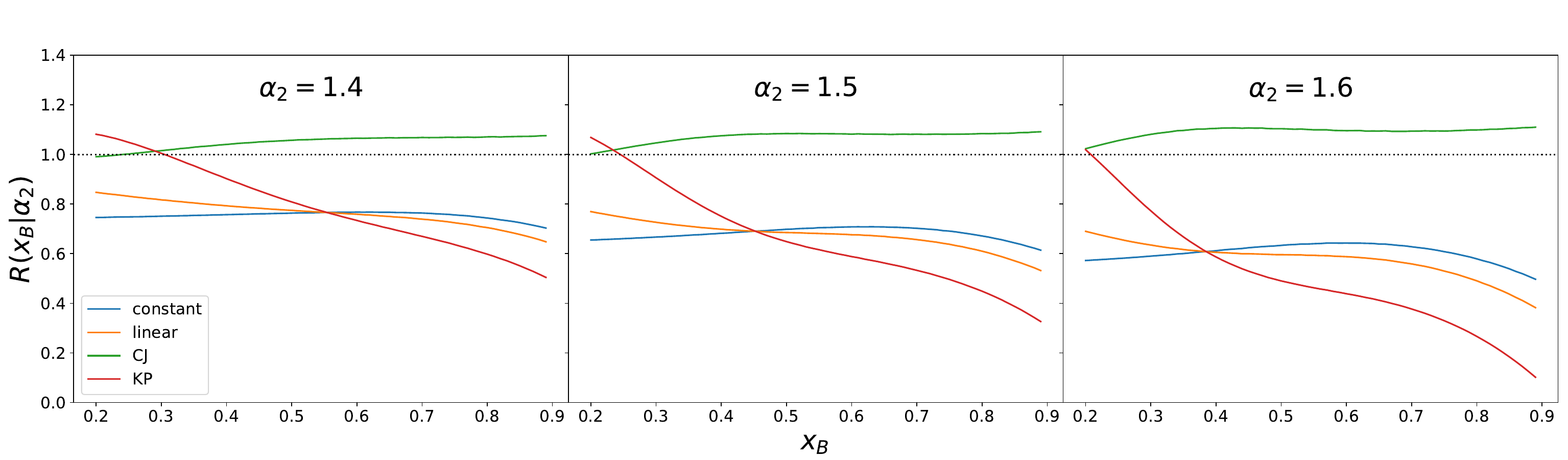}
    \caption{The results for the ratio of $F_2^A$ structure functions modified by the specified off-shell nucleon modification function models to the unmodified structure functions, shown against $x_B$.}
    \label{fig:single_xdep}
\end{figure}

First, we consider the effect of virtuality-dependent modification by calculating the ratio
\begin{equation}
R(x_B,Q^2,\alpha_2,\vec{p}_2^\perp)=\frac{F_{2,\text{modified}}^{A,\text{tag}}(x_B,Q^2,\alpha_2,\vec{p}_2^\perp)}{F_{2,\text{unmodified}}^A(x_B,Q^2,\alpha_2,\vec{p}_2^\perp)},
\end{equation}
of modified to unmodified tagged structure functions. We have chosen to evaluate this ratio at a fixed $Q^2=2$~GeV$^2/c^2$, a typical value for the coverage of CLAS12, given a 6~GeV electron beam. We have also chosen to evaluate this ratio at $\lvert \vec{p}_2^\perp \rvert = 0$, which is typical for the coverage of BAND detector. For simplicity, we omit these in our notation:
\begin{equation}
    R(x_B,\alpha_2)= R(x_B,Q^2=2\text{~GeV}^2/c^2,\alpha_2,\vec{p}_2^\perp=0).
\end{equation}
Figures \ref{fig:single_alphadep} and \ref{fig:single_xdep} show the tagged structure function ratio, $R(x_B,\alpha_2)$ as functions of $\alpha_2$ and $x_B$, respectively. We performed this calculation at three values of $x_B$ for the $\alpha_2$-dependence figures and vice versa. In both figures, one can see that these differing models yield significantly varied results, and that this variance increases with $\alpha_2$. 

The predictions of models we considered, all
of whom were tuned to existing data, range from below 0.4 to above 1 at a $\alpha_2=1.6$ and $x_B=0.7$. The KP model is most similar to the constant- and linear-in-$x$ models, though it has different behavior as a function of $x_B$. Surprisingly, the virtuality-dependent modification need not lead to a reduction in the structure function. The $CJ$ model predicts that the structure function will cause an increase from the no-modification assumption. We also note that even at $x_B=0.3$, a where medium-modification effects are typically thought of as being small, the different models predict 20--40\% variation in the tagged structure function, depending on $\alpha_2$.

We expect that the calculations will become less accurate with decreasing $\alpha_2$. Generalized Contact Formalism relies on a separation between the high-momentum scale of the relative momentum of nucleons within a correlated pair, from the low-momentum scale of typical momenta of nucleons with the nucleus. For pairs with a relative momentum comparable to the Fermi-momentum and below, these scales cease to be separated. For $p_2^\perp=0$, this occurs at roughly $\alpha_2=1.3$, which we have marked with a dashed line in Fig.~\ref{fig:single_alphadep}. We have analytically continued the calculation below $\alpha_2=1.3$, but cannot say with any confidence whether the structures observed in Fig.~\ref{fig:single_alphadep} are real, or an artifact of using GCF in an improper regime.

\begin{figure}[htpb!]
    \centering
    \includegraphics[width=4.5 in]{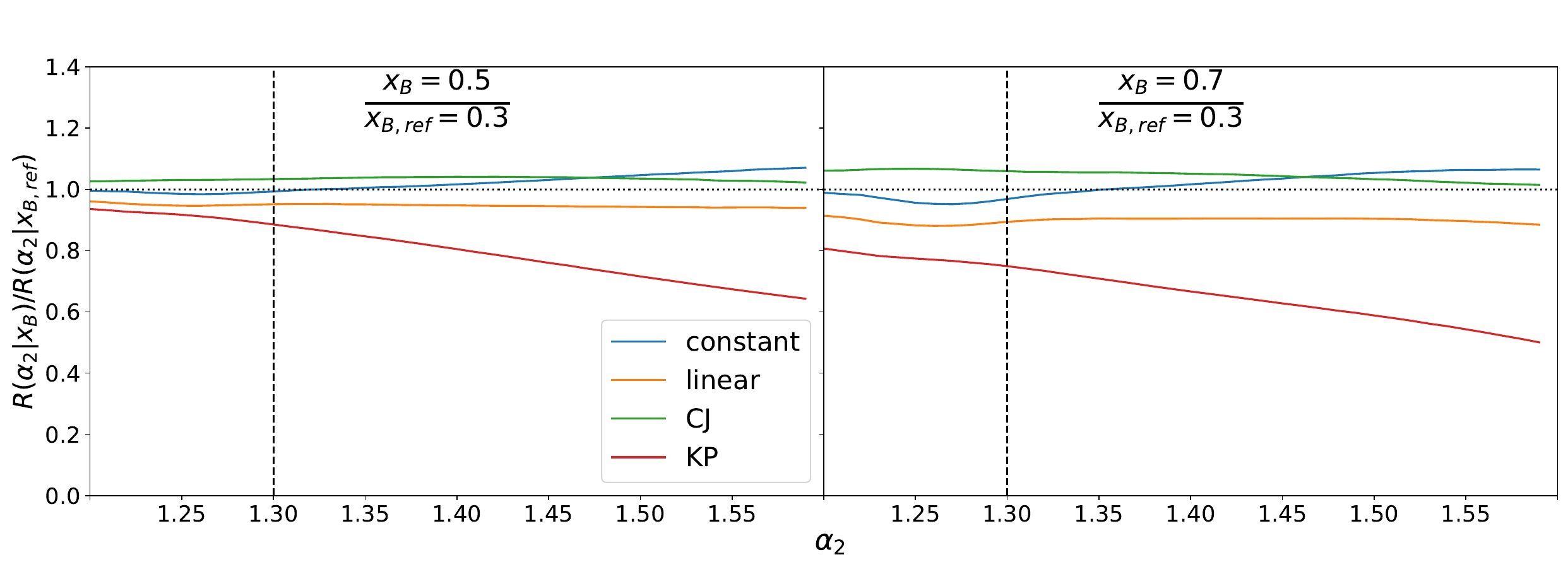}
    \caption{The results for the double ratio of the structure function at different $x_B$ values for the different off-shell nucleon modification function models}
    \label{fig:double_alphadep}
\end{figure}

\begin{figure}[htpb!]
    \centering
    \includegraphics[width=4.5 in]{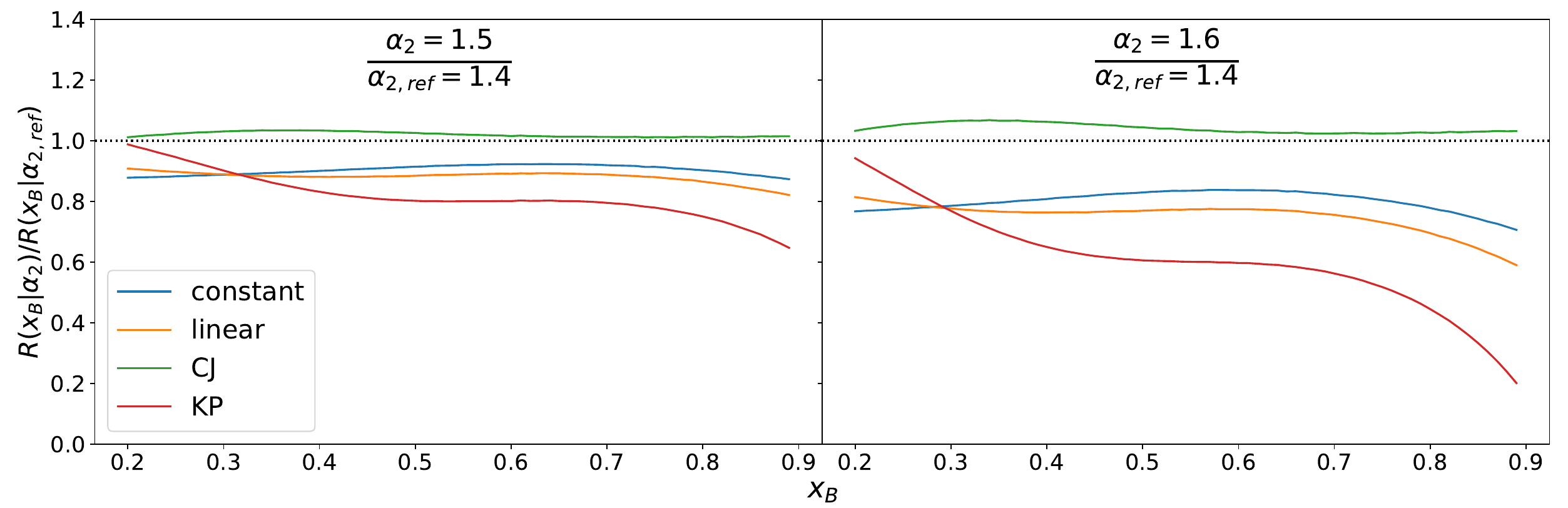}
    \caption{The results for the double ratio of the structure function at different $\alpha_2$ values for the different off-shell nucleon modification function models}
    \label{fig:double_xdep}
\end{figure}

In experiments, systematic effects can often be suppressed by forming ratios. For this reason, we also consider the double ratio $R(x_B,\alpha_2)/R(x_\text{B,ref},\alpha_2)$ as a function of $\alpha_2$ (Fig.~\ref{fig:double_alphadep}) and $R(x,\alpha_2)/R(x,\alpha_{2,\text{ref.}})$ 
as a function of $x_B$ (Fig.~\ref{fig:double_xdep}). The reference value $x_{B,\text{ref}}=0.3$ was chosen since structure function ratios $2F_2^A/AF_2^d$ are generally close to unity at $x_B=0.3$. The reference value $\alpha_{2,\text{ref}}=1.4$ was chosen as the smallest of the three values we considered (and hence corresponds to the least-virtual spectators).

As in the single ratios, the double ratios in Figs.~\ref{fig:double_alphadep} and \ref{fig:double_xdep} show a large sensitivity to the exact form of the virtuality-dependent modification. At $\alpha_2=1.6$ and $x_B=0.7$, the models have a spread from $\sim0.6$ to greater than $1$ , with KP and CJ being  most and least extreme, respectively. Furthermore, the spread between predictions increases with $x_B$ and $\alpha_2$. A measurement of $F_2^{^4\text{He,tag}}(x_B,\alpha_2)/F_2^{^4\text{He,tag}}(x_{B,\text{ref}},\alpha_{2,\text{ref}})$, normalized by a calculation of $F_{2,\text{unmodified}}^{^4\text{He,tag}}(x_B,\alpha_2)/F_{2,\text{unmodified}}^{^4\text{He,tag}}(x_{B,\text{ref}},\alpha_{2,\text{ref}})$ could provide new constraints on the possible form that virtuality-dependent modification can take. Even a measurement with 10\% uncertainty would be meaningful.

\section{Conclusions}

We describe a formalism for calculating the spectator-tagged structure function for scattering from a nucleon participating in a short-range correlation, using Generalized Contact Formalism to specify the two-nucleon light-cone density. We have performed calculations for helium-4 in kinematics relevant for the CLAS12-SRC Experiment and find large sensitivity to the assumed virtuality-dependent modification. Data from the CLAS12-SRC Experiment may be able to provide constraints on the nature of this modification. The method developed here can be naturally extended to other nuclei for which GCF parameters are known with reasonable uncertainty, including carbon-12 and calcium-40.

This approach has some specific limitations that would need to be addressed in order to make contact with experimental data. First and foremost, 
the validity of Generalized Contact Formalism can only be expected when there is scale separation between the high momentum of the correlated nucleons and the Fermi momentum, i.e., at large $\alpha_2$.
Extending the predictive range of these calculations to lower values of $\alpha_2$ will require a more complete description of
$\rho(\alpha_1,\vec{p}_1^\perp, \alpha_2, \vec{p}_2^\perp)$ that can cover the mean-field regime, the SRC regime, and
the transition between them. This is a more comprehensive description of the nucleus than a simple single-nucleon spectral function
of the kind employed in, for example, Refs.~\cite{Benhar:1997vy,Benhar:2000wi,Segarra:2020plg}. As a first step, one might 
consider treating the mean-field and SRC regimes separately and combining them in an ad hoc fashion. A more sophisticated route would be to develop a two-nucleon light-cone density using ab initio techniques such as variational Monte Carlo. We are not aware of any such calculations having been performed yet. In any case, care must be taken to ensure that the resulting $\rho(\alpha_1,\vec{p}_1^\perp, \alpha_2, \vec{p}_2^\perp)$ satisfies the baryon and momentum sum rules.

A second limitation is that our treatment only considers the $F_2$ structure function and does not make predictions for a cross section, which will have contributions from both $F_2$ and $F_1$, or, equivalently, the longitudinal cross section $\sigma_L$ and the transverse cross section $\sigma_T$. Our formalism cannot make any statement about the ratio $\sigma_L/\sigma_T$, much less about the possibility of nucleus dependence. Additional theoretical input on this question will be needed, and its uncertainty will be a critical uncertainty for the interpretation of any experimental measurement.

Lastly, our prescription considers only initial-state modification of the nuclear structure function; the interpretation of any measurement will also need to consider the effects of final-state interactions (FSIs). Without a full treatment of FSIs, such as in Refs.~\cite{Strikman:2017koc, Cosyn:2017ekf}, it will not be possible to interpret the results in terms of bound nucleon modification. One experimental handle could be to measure tagged DIS across multiple light nuclei, over which FSI effects can be expected to increase with nuclear size.

Spectator-tagged DIS from nuclei may offer additional insight into the mechanisms driving the EMC Effect and the role played by short-range correlations. The data from the CLAS12-SRC Experiment, specifically using the BAND detector for spectator neutron tagging, offer an excellent opportunity for such studies. 

\backmatter

\bmhead{Acknowledgments}

This work was supported by the US Department of Energy Office of Science, Office of Nuclear Physics, under contract no. DE-SC0016583. 
The authors are also grateful to J.~R.~Pybus, and T.~Kutz for helpful discussions. 

\bibliography{sn-bibliography}

\end{document}